\documentclass[12pt]{article}
\usepackage{amssymb}
\usepackage{amsmath}

\begin{document}

\title{A NONEQUILIBRIUM THERMODYNAMIC APPROACH TO GENERALIZED STATISTICS FOR
BROWNIAN MOTION}
\author{I. Santamar\'{\i}a-Holek\thanks{%
Fellow of SNI, Mexico. } \\
%EndAName
Facultad de Ciencias \\
Universidad Nacional Aut\'{o}noma de M\'{e}xico\\
Circuito Exterior, Cd. Universitaria. 04510 M\'{e}xico, D.F. M\'{e}xico\\
and \and R. F. Rodr\'{\i}guez\thanks{%
Fellow of SNI, Mexico. Correspondence author. E-Mail:
zepeda@fenix.ifisicacu.unam.mx.} \\
%EndAName
Instituto de F\'{\i}sica\\
Universidad Nacional Aut\'{o}noma de M\'{e}xico\\
Apdo. Postal 20-364, 01000\\
M\'{e}xico, D. F. M\'{e}xico.\\
Also at FENOMEC.}
\date{}
\maketitle

\begin{abstract}
We analyze the dynamics of a Brownian gas in contact with a heat bath in
which large temperature fluctuations occur. There are two distinct time
scales present, one describes the decay of the fluctuations in the
temperature and the other one is associated with the establishment of local
equilibrium. Although the gas has reached local equilibrium,
there exist large fluctuations in an intensive parameter (temperature) which
break the thermodynamic equilibrium with the heat bath. Thus the decay of
the fluctuations in the intensive parameter is larger than the
characteristic time for the establishment of local equilibrium. We show that
the dynamics of such large and intensive fluctuations may be described by
adopting a nonequilibrium thermodynamics approach with an adequate
formulation of local equilibrium. A coarsening procedure is then used to
contract the space of mesoscopic variables needed to describe the dynamics
of the gas and the extensive character of the description is lost. This
procedure allows us to derive an effective Maxwell-Boltzmann factor (EMBF)
for the Brownian gas, as has been recently proposed in the literature \cite%
{beck1}. Furthermore, we use this local equilibrium distribution and an
entropy functional to derive a nonequilibrium probability distribution and a
hydrodynamic description for the Brownian gas which contains fluctuating
transport coefficients. The ensuing description is nonextensive and our
analysis shows that the coarse-graining procedure is responsible for the
nonextensivity property.
\end{abstract}

\baselineskip=.5cm

Author Keywords: Nonequilibrium systems, Fluctuations of temperature,
Effective Boltzmann factor, Tsallis statistics


\newpage

\section{Introduction}

Complex systems where long range interactions, large external gradients or
large fluctuations of their intensive variables are present, show peculiar
behaviors which have attracted a good deal of attention in recent years \cite%
{sengers,tremblay,pagona,rodriguez,boghosian}. The theoretical description
of these systems constitutes a nowadays challenge and is far beyond the
limit of applicability of the theory of fluctuations near equilibrium as
developed by Onsager and Machlup \cite{onsager} and subsequently generalized
by Fox and Uhlenbeck \cite{fox}. This is essentially due to the fact that
these theories are restricted to describe fluctuations of extensive
variables only.

Several theories have been recently proposed to describe the special
properties of simple and complex fluids in nonequilibrium stationary states 
\cite{sengers,tremblay,pagona,rodriguez,boghosian,santamaria1}. However,
theories describing physical situations where fluctuations of the intensive
variables, such as temperature or chemical potential among others, may be
important are scarce, \cite{tsallis0,beck2}. These issues have been mostly
investigated within the framework of the so called nonextensive statistical
mechanics (NESM) \cite{tsallis0}, \cite{abe}, which according to Beck and
Cohen \cite{beck1}, is only one of \ the possible more general statistics
that can deal with the physical situations mentioned above. NESM proposes
that for systems with a sufficiently complex dynamics, its necessary to
introduce a more general statistics than the usual Boltzmann-Gibbs
description. Such statistics have been discussed for a variety of systems 
\cite{beck1} and have led to the introduction of an effective Boltzmann
factor (EMBF) which reduces to the ordinary one in the appropriate limit.

These concepts have been explored in more detail within the model of
Brownian motion, for which Beck and Cohen have postulated a modification of
the Boltzmann factor whose explicit form depends on the statistics of the
intensive parameters fluctuations. They show that if it is assumed that the
inverse temperature of ordinary statistical mechanics fluctuates with the
Gamma distribution, then the ensuing Brownian dynamics can be correctly
described by Tsallis statistics, \cite{beck1,beck2}. One of the main
purposes of this paper is to use this same model of nonequilibrium Brownian
motion to examine how the generalization of the Boltzmann factor may be
justified in terms of \ well established concepts used in irreversible
thermodynamics, such as Einstein's formula for the probability distribution
and a Kullback-like entropy functional \cite{schlogl, wehrl,degroot,mazur}.
By adopting this point of view, we show that the modified Boltzmann's factor
has its origin in the contraction of the space of mesoscopic variables that
describe the dynamics of the system, and that this contraction is
responsible for the non-extensive character of the new description.

Usually, the dynamics of a Brownian gas is analyzed by taking into account
only the equilibrium fluctuations of the extensive variables of the heat
bath, and by assuming thermodynamic equilibrium between the Brownian gas and
the bath. Since the characteristic decaying time of these fluctuations is
short as compared with the time characterizing the establishment of local
equilibrium, they become independent from those of the velocity of the
Brownian particles which, in this case, are well described by the
Maxwell-Boltzmann factor \cite{landau sp}. However, a more general case
occurs when intensive parameters (temperature) may fluctuate and are large
enough to break its thermodynamic equilibrium with the heat bath. In this
case, the Brownian gas can be assumed to be in local thermodynamic
equilibrium, but not with the entire heat bath. In this situation the work
performed on the gas will be different from that corresponding to the
equilibrium case. These conditions can be met when the system is constrained
to be far away from equilibrium and the time scale and wavelength of the
intensive fluctuations are larger than those characterizing the
establishment of local equilibrium. As a consequence, the probability
distribution for the fluctuating velocities of the Brownian particles
depends, in general, on the distribution of the temperature fluctuations and
is no longer described by the usual Maxwell-Boltzmann factor.

The paper is organized as follows. In Sec. \textbf{II} we define the system
and obtain a probability distribution for the velocities of the Brownian
particles that incorporates the effects of nonequilibrium temperature
fluctuations; from it we derive the effective Maxwell-Boltzmann factor. Then
in Sec. \textbf{III} we derive a Fokker-Planck equation describing the
relaxation mesoscopic dynamics of the Brownian gas and we use it to derive a
hydrodynamic description with fluctuating transport coefficients. A
generalized Onsager-Machlup formalism is presented in section \textbf{IV}
and the last section \textbf{V}\ is devoted to discuss our main results.

\section{Model and Basic Equations}

As mentioned above, the dynamics of a diluted gas of Brownian particles in
contact with a heat bath in the presence of two distinct time scales, one
related to the decay of the fluctuations and the other to the establishment
of local equilibrium, can be interpreted by assuming that
each Brownian particle is embedded in a system in contact with a heat bath
at temperature $T_{B}(t)$. We want to analyze the dynamics of the Brownian
particle when nonequilibrium temperature fluctuations $\delta T(t)$ occur in the system. In particular, we will assume that the characteristic
length and time scales of these fluctuations are sufficiently larger than
those characterizing the establishment of local equilibrium. The relation
between the system and heat bath temperatures is then $T(t)=T_{B}(t)+\delta
T(t)$.

When a fluctuation occurs, the system only performs the amount of work $W$
on the Brownian particle. This modifies its kinetic energy by $W/m=\delta
e_{kin}=-\frac{\overrightarrow{u}^{2}}{2}$, where $m$ is the mass of the
particle and the sign indicates that the system performs the work. Moreover,
the system exerts the work $p\delta V$ and transfers the amount of heat $%
T\delta S$ to the bath. Assuming that the variation of the total entropy of
the system plus bath is $\delta S_{T}=\delta S+\delta S_{B}$, and by using
the Gibbs equation, after dividing by $m$ we obtain the relation \cite%
{landau sp,degroot} 
\begin{equation}
\delta s_{T}=\frac{1}{T(t)}\left[ \delta e_{kin}+\delta e+p\delta \rho ^{-1}%
\right] +\frac{1}{T_{B}(t)}\left[ \delta e_{B}+p_{B}\delta \rho _{B}^{-1}%
\right] ,  \label{st1}
\end{equation}%
where $\delta e$ and $\delta e_{B}$ are the variations of the internal
energy of the system and the bath, $p$ and $p_{B}$ the system and bath
pressures, and $\rho $ and $\rho _{B}$ the corresponding specific volumes.

Eq. (\ref{st1}) can be used to obtain the minimum amount of work performed
by the system to modify the state of the Brownian particle \cite{landau sp}.
This situation occurs when the processes taking place between the system and
the heat bath are reversible. Since in that case $\delta e_{B}=-\delta e$
and $\delta \rho _{B}^{-1}=-\delta \rho ^{-1}$, and given that $\delta
T=T-T_{B}$, then Eq. (\ref{st1}) becomes 
\begin{equation}
\delta s_{T}=\frac{1}{T(t)}\delta e_{kin}-\frac{1}{T\,T_{B}}\delta T\delta
e+\left( \frac{p}{T}-\frac{p_{B}}{T_{B}}\right) \delta \rho ^{-1},
\label{st2}
\end{equation}%
which is in agreement with the Le Chatelier-Braun principle \cite{landau
sp,degroot}. In order to write equation (\ref{st2}) in a more convenient
form, we expand $\delta e(T,\rho )$ up to first order in $\delta T$ and $%
\delta \rho ^{-1}$. After using the relation $\left( \frac{\partial e}{%
\partial \rho }\right) _{T}=-p+T(t)\left( \frac{\partial p}{\partial T}%
\right) _{\rho }$ we arrive at 
\begin{equation}
\delta s_{T}=\frac{1}{T(t)}\delta e_{kin}-\frac{1}{T\,T_{B}}\left( \frac{%
\partial e}{\partial T}\right) _{\rho }\left( \delta T\right) ^{2}+\left[ 
\frac{p}{T\,T_{B}}-\frac{1}{T_{B}}\left( \frac{\partial p}{\partial T}%
\right) _{\rho }\right] \delta T\delta \rho ^{-1}+\left( \frac{p}{T}-\frac{%
p_{B}}{T_{B}}\right) \delta \rho ^{-1},  \label{st3}
\end{equation}%
which can still be simplified if we use again the relation $T=T_{B}+\delta T$
in order to make the approximations 
\begin{equation}
\frac{p}{T\,T_{B}}\simeq \frac{p}{T_{B}^{2}}\left( 1-\frac{\delta T}{T_{B}}%
\right) ,\,\,\,\,\,\,\,\,\,\,\,\,\,\,\,\,\frac{p}{T}-\frac{p_{B}}{T_{B}}%
\simeq \frac{\delta p}{T_{B}}-\frac{p}{T_{B}}\frac{\delta T}{T_{B}},
\label{aprox1}
\end{equation}%
where $\delta p\equiv p-p_{B}$. After substitution of Eq. (\ref{aprox1})
into Eq. (\ref{st3}), keeping terms up to second order in the fluctuations
and using the expansion $\delta p\simeq \left( \frac{\partial p}{\partial
\rho }\right) _{T}\delta \rho +\left( \frac{\partial p}{\partial T}\right)
_{\rho }\delta T$, we arrive at the following expression for the variation
of the total entropy $\delta s_{T}$ 
\begin{equation}
\delta s_{T}=\frac{1}{T(t)}\delta e_{kin}-\frac{1}{T\,T_{B}}\left( \frac{%
\partial e}{\partial T}\right) _{\rho }\left( \delta T\right) ^{2}-\frac{1}{%
T_{B}}\left( \frac{\partial p}{\partial \rho }\right) _{\rho }\rho
^{-2}\left( \delta \rho \right) ^{2}.  \label{st4}
\end{equation}

However, since our interest in this paper is to analyze the effects of the
large temperature fluctuations on the dynamics of the Brownian gas, we shall
adopt a statistical description point of view and assume that the
probability $P_{le}$ of occurrence of a fluctuation in the composite system
may be expressed through Einstein's formula \cite{landau sp} 
\begin{equation}
P_{le}=P_{0}e^{\frac{m\delta s_{T}}{k_{B}}},  \label{Einstein}
\end{equation}%
where $k_{B}$ is Boltzmann's constant and $P_{0}$ normalizes $P_{le}$. Thus,
by substituting Eq. (\ref{st4}) into (\ref{Einstein}) and using $\delta
e_{kin}=-\frac{\overrightarrow{u}^{2}}{2}$, $P_{le}$ can be rewritten in the
more convenient form 
\begin{equation}
P_{le}=P_{0}e^{-\frac{m}{2k_{B}T_{B}(t)}h(\delta T)\left\{ \overrightarrow{u}%
^{2}+\frac{2}{T_{B}}\left( \frac{\partial e}{\partial T}\right) _{\rho
}\left( \delta T\right) ^{2}+h^{-1}(\delta T)\left( \frac{\partial p}{%
\partial \rho }\right) _{\rho }\rho ^{-2}\left( \delta \rho \right)
^{2}\right\} },  \label{PRmin}
\end{equation}%
which is the probability that a fluctuation occurs. To arrive at Eq. (\ref%
{PRmin}) we have expanded $\frac{1}{T(t)}\simeq \frac{1}{T_{B}(t)}h(\delta T)
$ with $h(\delta T)\simeq \left[ 1-\frac{\delta T}{T_{B}(t)}-\left( \frac{%
\delta T}{T_{B}(t)}\right) ^{2}\right] $. This last equation allows us to
analyze how the distribution function accounting for the fluctuations of the
velocities of the Brownian particle becomes modified in the presence of
large temperature fluctuations $\delta T$ \ occuring in the system.
Integrating Eq. (\ref{PRmin}) over $\delta T$ leads to the effective
Maxwell-Boltzmann factor $B_{le}$ 
\begin{equation}
B_{le}=P_{0}\int e^{-\frac{m}{2k_{B}T_{B}(t)}h(\delta T)\left[ 
\overrightarrow{u}^{2}+2\left( \frac{\partial e}{\partial T}\right) _{\rho }%
\frac{\left( \delta T\right) ^{2}}{T_{B}(t)}\right] }d(\delta T),
\label{PRmin4}
\end{equation}%
where we have assumed a constant volume processes for simplicity. This
probability distribution is an effective or modified Boltzmann factor (EMBF)
for the nonequilibrium Brownian system and it contains corrections arising
from a non-Gaussian distribution. However, in the usual case where system
and bath are in thermal equilibrium, $h(\delta T)\rightarrow 1$, and the
temperature fluctuations become independent of velocity fluctuations, as
expected \cite{landau sp}.

\subsection{Connection with superstatistics}

Equation (\ref{PRmin4}) implies that when system and bath are not in thermal
equilibrium, temperature fluctuations modify the local equilibrium
distribution of the Brownian particle in the sense discussed in Ref. \cite%
{beck1}. To illustrate this point more explicitly, let us rewrite (\ref%
{PRmin4}) as 
\begin{equation}
B_{le}(e_{B})=\int f\left( \delta T\right) e^{-h(\delta T)e_{B}}d(\delta
T)=\langle e^{-h(\delta T)e_{B}}\rangle _{f},  \label{PRmin5}
\end{equation}%
where $e_{B}\equiv \frac{m\,\overrightarrow{u}^{2}}{2k_{B}T_{B}(t)}$ and $%
\langle ..\rangle _{f}$ denotes the average over the distribution function
of the temperature fluctuations $f\left( \delta T\right) $ given by 
\begin{equation}
f\left( \delta T\right) =P_{0}e^{-\frac{m\,h(\delta T)}{k_{B}T_{B}(t)}\left[
\left( \frac{\partial e}{\partial T}\right) _{\rho }\frac{\left( \delta
T\right) ^{2}}{T_{B}(t)}\right] }.  \label{fbeta}
\end{equation}%
If we now follow a procedure similar to that sketched in Ref. \cite{beck1},
we may write the relation 
\begin{equation}
B_{le}(e_{B})=e^{-\langle h\rangle _{f}\,e_{B}}\langle e^{-\left( h-\langle
h\rangle _{f}\right) \,e_{B}}\rangle _{f},  \label{Blebracket1}
\end{equation}%
where $\langle h\rangle _{f}$ is the average of $h(\delta T)$ over the
distribution $f\left( \delta T\right) $,(\ref{fbeta}). By expanding the
exponential within the angular bracket up to second order in its argument
and after evaluating the average we obtain 
\begin{equation}
B_{le}(e_{B})\simeq e^{-\langle h\rangle _{f}e_{B}}\left[ 1-\frac{1}{2}%
\Sigma ^{2}e_{B}^{2}\right] ,  \label{BleTsallis}
\end{equation}%
where we have introduced the abbreviation $\Sigma ^{2}\equiv \frac{\langle
\delta T^{2}\rangle _{f}}{T_{B}^{2}}$. It should be stressed that this
result was obtained from a nonequilibrium thermodynamic analysis and
expresses the effective Maxwell-Boltzmann factor in an universal form which
only depends on the second moment, $\langle \delta T^{2}\rangle _{f}$, of
the temperature fluctuations. This corresponds to the low-E universal
behavior discussed in Ref. \cite{beck1}, where this universality allows for
the definition of a universal parameter for any superstatistics and not only
for Tsallis statistics.

\section{Hydrodynamics with fluctuating transport coefficients}

\bigskip As mentioned in the introduction, fluctuations in intensive
parameters such as the temperature, can not be described by the usual
Onsager-Machlup theory, since it is devised to describe fluctuations in the
extensive parameters only \cite{onsager}. For this reason, here we will
describe the dynamics of a Brownian particle in the presence of large local
temperature fluctuations $\delta T(t)$ in its $(\vec{r},\vec{u})$-space,
where $\vec{r}$ and $\vec{u}$ stand for the position and the velocity of a
Brownian particle. More precisely, the state of the particle is defined not
only by $\vec{r}$ and $\vec{u}$, but also by the value of the parameter $%
\delta T(t)$  entering in the single particle
nonequilibrium probability distribution in the form $P(\vec{r},\vec{u};t,\delta T)$. To accomplish the description, note that at the
mesoscale the nonequilibrium probability distribution $P$ obeys the continuity equation 
\begin{equation}
\frac{\partial }{\partial t}P(\vec{r},\vec{u};t,\delta T)+\nabla \cdot \left[
\vec{u}P\right] =-\frac{\partial }{\partial \vec{u}}\cdot \left( P\vec{v}_{%
\vec{u}}\right) ,  \label{continuidadP}
\end{equation}%
where $\nabla \equiv \frac{\partial }{\partial \vec{r}}$, $\vec{v}_{\vec{u}}(%
\vec{r},\vec{u};t,\delta T)$ is the corresponding streaming velocity in $%
\vec{u}$ -space. In this case, the explicit form of Eq. (\ref{continuidadP})
can be constructed by using Eq. (\ref{PRmin}) and by using the following
functional for the local nonequilibrium Kullback-like entropy $s(\vec{r},t)$ 
\cite{schlogl,wehrl,mazur,kampen,aperez,PNAS,santamaria1} 
\begin{equation}
s(\vec{r},t)=-k_{B}\int P(\vec{r},\vec{u};t,\delta T)\ln \frac{P}{P_{le}}%
\,\,\,d\vec{u}\,,  \label{p. gibbs}
\end{equation}%
where $P_{le}$ characterizes a (local) equilibrium reference state, and is
given by Eq. (\ref{PRmin}) for $\delta \rho =0$. It is important to stress
that the entropy functional (\ref{p. gibbs}) will be used here as an
irreversibility criterion that reduces to the Gibbs entropy in the case of
equilibrium \cite{wehrl,mazur,kampen}.

By taking the time derivative of Eq. (\ref{p. gibbs}) and by substitution of
the result into Eq. (\ref{continuidadP}), one may calculate the entropy
production $\sigma (\vec{r},t)$ of the Brownian gas through its natural
evolution in time. After integrating by parts and by assuming that the
currents vanish at the boundaries, one arrives at 
\begin{equation}
\frac{\partial s}{\partial t}+\nabla \cdot \vec{J}_{s}=-k_{B}\int P\vec{v}_{%
\vec{u}}\cdot \frac{\partial }{\partial \vec{u}}\ln \frac{P}{P_{le}}\,\,\,d%
\vec{u}\,,  \label{sigma}
\end{equation}%
where we have defined the entropy flow by 
\begin{equation}
\vec{J}_{s}\equiv -k_{B}\int P\vec{u}\left( \ln \frac{P}{P_{le}}-1\right)
\,\,\,d\vec{u}\,.  \label{mu}
\end{equation}%
For convenience we have assumed that the average temperature of the bath is
constant. The right hand side of Eq. (\ref{sigma}) is the entropy production
in the space of variables determining the state of the gas in the mesoscale, 
\cite{mazur}, \cite{aperez}.

Given the entropy production in (\ref{sigma}), one may impose the condition
of a (locally) positive entropy production in the $(\vec{r},\vec{u})$-space
and by following a scheme similar to that of nonequilibrium thermodynamics,
in first approximation formulate linear laws between the current $P\vec{v}_{%
\vec{u}}$ and its conjugated force $\frac{\partial }{\partial \vec{u}}\ln 
\frac{P}{P_{le}}$, \cite{degroot}, \cite{PNAS}. After taking into account Eq. (\ref{PRmin}), the linear law takes the form
\begin{equation}
P\vec{v}_{\vec{u}}=\beta h(\delta T)\vec{u}P+\frac{k_{B}T_{B}}{m}\beta \frac{%
\partial P}{\partial \vec{u}}  \label{linearlaw1}
\end{equation}%
where $\beta $ is the corresponding coupling coefficient which has been
assumed to be a scalar function for simplicity. Near equilibrium, this
coefficient must reduce to the corresponding Onsager coefficient. Now, by
substituting Eq. (\ref{linearlaw1}) into (\ref{continuidadP}), we finally
arrive at the following Fokker-Planck equation 
\begin{equation}
\frac{\partial }{\partial t}P(\vec{r},\vec{u};t,\delta T)+\nabla \cdot \left[
\vec{u}P\right] =\frac{\partial }{\partial \vec{u}}\cdot \left( \beta
h(\delta T)\vec{u}P+\frac{k_{B}T_{B}}{m}\beta \frac{\partial P}{\partial 
\vec{u}}\right) ,\,\,\,\,\,\,\,\,\,\,\,  \label{FP-FH}
\end{equation}%
which describes the evolution in time of the nonequilibrium probability
distribution of the Brownian particle. It is important to point out that the
factor $h(\delta T)$ breaks the equilibrium form of the
fluctuation-dissipation theorem (FDT), thus implying that in the presence of
large temperature fluctuations the usual form of the FDT is not valid. This
result is due to the lack of thermal equilibrium between the system and the
heat bath.

Once obtained the Fokker-Planck equation (\ref{FP-FH}) it is now possible to
derive an hydrodynamic-like hierarchy of equations for the moments of the
probability distribution function $P(\vec{r},\vec{u};t,\delta T)$. As shown
below, these equations constitute a fluctuating hydrodynamic description
since they contain fluctuating transport coefficients. The evolution
equation for the mass density field $\rho (\vec{r};t,\delta T)=m\int Pd\vec{u%
}$ is simply obtained by directly integrating Eq. (\ref{FP-FH}) over the
velocity $\vec{u}$, yielding 
\begin{equation}
\frac{\partial }{\partial t}\rho (\vec{r};t,\delta T)=-\nabla \cdot \left[
\rho \vec{v}(\vec{r};t,\delta T)\right] .  \label{mass}
\end{equation}%
In a similar form, the evolution equation for the momentum field $\rho \vec{v%
}(\vec{r};t,\delta T)\equiv m\int \vec{u}Pd\vec{u}$ can be shown to be 
\begin{equation}
\rho \frac{d}{dt}\vec{v}(\vec{r};t,\delta T)+\nabla \cdot \!\!\,\,\vec{\vec{%
\mathrm{P}}}^{k}\!\!\,\,=-\rho \beta ^{\ast }\cdot \vec{v},
\label{evol-momentum}
\end{equation}%
where $\beta ^{\ast }=\beta h\left( \delta T(t)\right) $ is a fluctuating
friction coefficient. Here we have introduced the kinetic part of the
pressure tensor $\,\vec{\vec{\mathrm{P}}}^{k}\!\!$ 
\begin{equation}
\!\!\,\,\vec{\vec{\mathrm{P}}}^{k}\!\!\,\,(\vec{r};t,\delta T)=m\int (\vec{u}%
-\vec{v})(\vec{u}-\vec{v})Pd\vec{u},  \label{Puu}
\end{equation}%
which obeys the following relaxation equation 
\begin{equation}
\frac{d}{dt}\!\!\,\,\vec{\vec{\mathrm{P}}}^{k}\!\!\,\,+2\left[ \!\!\,\,\vec{%
\vec{\mathrm{P}}}^{k}\!\!\,\,\cdot \left( \beta ^{\ast }\vec{\vec{1}}+\nabla 
\vec{v}+\frac{1}{2}(\nabla \cdot \vec{v})\vec{\vec{1}}\right) \right] ^{s}=%
\frac{2k_{B}T_{B}}{m}\beta \rho \vec{\vec{1}},  \label{evolution-P}
\end{equation}%
obtained by taking the time derivative of Eq. (\ref{Puu}) and using (\ref%
{FP-FH}). Here $\vec{\vec{\mathrm{1}}}$ is the unit tensor, the superscript $%
s$ denotes the symmetric part of a tensor and we have neglected the
contribution of the third and higher order moments (see Ref. \cite{nosotros}%
). Eqs. (\ref{mass})-(\ref{evolution-P}) constitute a generalized
fluctuating hydrodynamic equations of the Brownian phase when large
temperature fluctuations drive the system out of equilibrium, breaking its
thermal equilibrium with the bath \cite{zwanzig}. It should be remarked that if
the usual internal thermal fluctuations associated with the term
proportional to $k_{B}T_{B}$ were to be taken into account, a random heat
and stress flows should be added to the elements of the pressure tensor $%
\,\,\!\!\vec{\vec{\mathrm{P}}}^{k}\!\!$. Then Eqs. (\ref{evol-momentum}) and
(\ref{evolution-P}) may be rewritten in the usual form of the Landau and
Lifshitz fluctuating hydrodynamic equations, \cite{landau 2}.

\subsection{The limit of long times}

At sufficiently long times $t\beta ^{\ast }>>1$, the description of the
Brownian motion can be performed in terms of a diffusion equation for $\rho (%
\vec{r};t,\delta T)$. This equation can be derived by neglecting the time
derivatives in Eqs. (\ref{evol-momentum}) and (\ref{evolution-P}). Then, Eq.
(\ref{evolution-P}) leads to the following constitutive relation for the
pressure tensor 
\begin{equation}
\!\!\,\,\vec{\vec{\mathrm{P}}}^{k}\!\!\,\,\simeq \frac{k_{B}T_{B}}{m}\rho
h^{-1}(\delta T)\vec{\vec{\mathrm{1}}},  \label{Pkfluc}
\end{equation}%
where, for simplicity, the Brownian contribution to the viscosity has been
neglected \cite{nosotros,santamaria1}. It should be noted in passing, that
precisely the same first two terms on the right hand side of Eq. (\ref{FP-FH}%
), which were responsible for the breaking of the fluctuation-dissipation
theorem in its usual form, are also responsible for the appearance of the
correction $h^{-1}(\delta T)$ on the right hand side of Eq. (\ref{Pkfluc}).
Substitution of (\ref{Pkfluc}) into Eq.(\ref{evol-momentum}) leads to the
following expression for the diffusion current $\rho \vec{v}$ 
\begin{equation}
\rho \vec{v}\simeq -D_{fluc}(\delta T)\nabla \rho ,  \label{HF-diffusion}
\end{equation}%
where the fluctuating diffusion coefficient has been defined as 
\begin{equation}
D_{fluc}(\delta T)\equiv \frac{k_{B}T_{B}}{m\beta }h^{-2}(\delta T).
\label{D_f}
\end{equation}%
From Eqs. (\ref{Pkfluc}) and (\ref{HF-diffusion}), one may expect that when
the fluctuations $\delta T$ are spatially non-homogeneous, then a
fluctuating heat flow contributes to the diffusion flow. These results imply
that in the long time limit under consideration and when large temperature
fluctuations occur in the system, the diffusion coefficient of the Brownian
particle is converted into a fluctuating quantity. The evolution equation
for the mass density $\rho (\vec{r};t,\delta T)$ in this limit is the
obtained by substituting Eq. (\ref{HF-diffusion}) into (\ref{mass}) yielding
the fluctuating Smoluchowski equation 
\begin{equation}
\frac{\partial }{\partial t}\rho (\vec{r};t,\delta T)=D_{fluc}(\delta
T)\nabla ^{2}\rho ,  \label{Smol-fluc}
\end{equation}%
which describes the evolution of the Brownian particle for long times. Eq. (%
\ref{Smol-fluc}) reinforces the fact that the fluctuation-dissipation
theorem in its usual form is no longer valid.

\section{Regression laws in Brownian dynamics}

We can also derive the regression laws for both, the conditional averages of
the velocity of the Brownian particles, and of the temperature fluctuations
over a given volume element $dV=d\vec{r}$. To accomplish this objective we
must take the average of the corresponding Fokker-Planck equation that also
incorporates the temperature fluctuation $\delta T$ , over $d\vec{r}$ and
then use the definitions of the conditional averages for the velocity and
for the temperature 
\begin{equation}
\overline{\vec{u}}^{0}(t)=\int \vec{u}P_{u}(\vec{u},\delta T;t)d\vec{u}%
\,d(\delta T)  \label{averageflucU}
\end{equation}%
and 
\begin{equation}
\overline{\delta T}^{0}(t)=\int \delta TP_{u}(\vec{u},\delta T;t)d\vec{u}%
\,d(\delta T).  \label{averageflucT}
\end{equation}%
The upper bar denotes a given initial state and 
\begin{equation}
P_{u}(\vec{u};t,\delta T)=\int P(\vec{r},\vec{u},\delta T;t)d\vec{r}.
\label{P reducida}
\end{equation}%
Using this definition, from the corresponding Fokker-Planck equation a time evolution equation for $P_{u}(\vec{u},\delta T;t)$ is readily obtained. If we insert the resulting equation into the time derivatives of Eqs. (\ref{averageflucU}) and (\ref%
{averageflucT}), after a partial integration similar to that of Sec. III, we arrive at the equations 
\begin{equation}
\frac{d\overline{\vec{u}}^{0}(t)}{dt}=-\tilde{\beta}(\overline{\delta T}%
^{0})\,\overline{\vec{u}}^{0}(t),  \label{regressionlawU}
\end{equation}%
and 
\begin{equation}
\frac{d\overline{\delta T}^{0}(t)}{dt}=-\epsilon \,\overline{X}_{\delta
T}^{0}(t),  \label{regressionlawT}
\end{equation}%
where $\epsilon$ is an Onsager coefficient and we have assumed that $\beta $ and $\epsilon $ are
independent of position. The nonlinear generalized force $\overline{X}_{\delta T}^{0}(t)$ has been
defined as 
\begin{equation}
\overline{X}_{\delta T}^{0}(t)=\int {\frac{\partial }{\partial \delta T}%
\left( \frac{h}{2}u^{2}+\left( \frac{\partial e}{\partial T(t)}\right)
_{\rho }\frac{\left( \delta T\right) ^{2}}{T_{B}}\right) }P_{u}d\vec{u}\,d{%
(\delta T)},  \label{XdeltaT}
\end{equation}%
and the modified friction coefficient $\tilde{\beta}$ through the relations 
\begin{eqnarray}
\tilde{\beta}(\overline{\delta T}^{0})\,\overline{\vec{u}}^{0}(t) &=&\int 
\vec{u}\beta h{(}\delta T{)}P_{u}d\vec{u}\,d(\delta
T)\,\,\,\,\,\,\,\,\,\,\,\,\,\,  \notag  \label{betatilde} \\
\,\,\,\,\,\,\,\,\,\,\,\,\,\, &=&\int \vec{u}{\left\{ \int \beta h(\delta
T)P_{u}d(\delta T{)\,}\right\} }d\vec{u}=\tilde{\beta}(\overline{\delta T}%
^{0})\int \vec{u}f(\vec{u},t)d\vec{u},
\end{eqnarray}%
where $f(\vec{u},t)=\int P_{u}(\vec{u},\delta T;t)d(\delta T)$ is a reduced
distribution function. By taking into account the explicit expression for $%
h(\delta T)$ and the fact that the average of $\delta T$ vanishes, one
obtains $\tilde{\beta}=\beta \left( 1-\frac{1}{T_{B}^{2}}(\overline{\delta T}%
^{0})^{2}\right) $. Note that the regression law for the temperature
fluctuations is a non-linear one, and that it is coupled with that for $%
\overline{\vec{u}}^{0}(t)$.

From Eqs. (\ref{regressionlawU}) and (\ref{regressionlawT}), it follows that
mean regression laws are obeyed by the time-dependent value of the
fluctuations occurring in the system after being averaged over an elemental
volume. It is interesting to note that in the present case, when external
factors induce temperature fluctuations, the average coefficients
characterizing the dissipation in the system become modified by a term
proportional to the square of the amplitude of the fluctuation. Eqs. (\ref%
{regressionlawU}) and (\ref{regressionlawT}) may be interpreted as the
generalization of the Onsager regression hypothesis in which the presence of
large temperature fluctuations introduce corrections to the dissipation
coefficients.

%If we assume that the temperature fluctuation gives rise to a fluctuation of the %potential energy of the Brownian particle, then the corresponding Smoluchowski %equation (\ref{Smol-fluc}) describing the evolution of the systems at large times in %terms of the mass density field $\rho=\rho\left(z,\delta T, t\right)$ can be written %as
%\begin{equation}
%\frac{\partial }{\partial t}\rho =\frac{\partial}{\partial z}\cdot \left( %\beta^{*}(\delta T) g \rho + D_{fluc}(\delta T) \frac{\partial \rho}{\partial z} %\right),
%\label{SmolBarometric}
%\end{equation}
%which has associated the following stochastic equation (with additive and %multiplicative noises) for $z(t)$
%\begin{equation}
%\frac{d}{dt} \overline{z(t)}^0=-\beta^{*}(\delta T) g + \vec{F}_R,  \label{Lang z}
%\end{equation}%
%where $\vec{F}_R$ is associated with the thermal fluctuation of the bath.
%\section{Applications?}

\section{DISCUSSION}

By using an irreversible thermodynamic approach we have derived an effective
Maxwell-Boltzmann factor (EMBF) for a gas of Brownian particles, when there
occur large fluctuations in the temperature of the heat bath in which they
are suspended. The derived EMBF has the same form as the one the proposed in
the literature \cite{beck1}. Our derivation was carried out by using
irreversible thermodynamic arguments of the type used to evaluate the
minimum amount of work that a system embedded into a heat bath needs to
perform in order to modify the state of an external body \cite{landau sp}.
The EMBF has been expressed in a form that relates it to the superstatistics
discussed in Ref. \cite{beck1}. The general point of view we have presented
here indicates that this EMBF had its origin in a contraction of the
mesoscopic space of variables necessary to completely describe the dynamics
of the system. In this case, the superstatistics present in this mesoscopic
stochastic variables have been eliminated by averaging over them. As a
consequence of this contraction the extensive character of the description
is lost.

A hydrodynamic description was derived from a Fokker-Planck equation for the
nonequilibrium distribution function of the Brownian system. The last
equation was obtained by using a Kullback-like entropy functional which
incorporates the complete distribution function (\ref{PRmin}) as a reference
state, and then averaging over the fluctuating velocities of the Brownian
particle. The resulting description can be considered as a generalized
fluctuating hydrodynamics scheme for the Brownian gas, that incorporates
fluctuations of the intensive variables through random transport
coefficients. Finally, by studying the appropriate space averages, we have
derived non-linear regression equations for the time dependent average of
the variables of the Brownian system.

The results we have obtained in this paper, show that the description of far
from equilibrium systems given in $(\vec{r},\vec{u},\delta T)$-space lead,
after contraction, to a non-extensive description. They also lead to a
generalized hydrodynamics with fluctuating transport coefficients. Actually,
since the transport coefficients depend on $\delta T$, the linear relation
between fluxes and forces in the $(\vec{r},\delta T)$-space is in general
lost, implying again a nonextensive description \cite{santamaria1}.

{\Large Acknowledgments}

We wish to acknowledge Profs. A. Robledo and J. M. Rub\'{\i} and Dr. A. P%
\'{e}rez-Madrid for critical reading of the manuscript and for clarifying
discussions. RFR acknowledges financial support from DGAPA-UNAM, Grant IN
112503.

\newpage

\end{document}